\documentclass[11pt]{article}
\usepackage{amsmath,amsbsy,amsfonts,amssymb,graphics,epsfig,float,mathrsfs,amsthm,braket}
\usepackage{ifsym}
\usepackage{psfrag}
\usepackage{color}
\usepackage[normalem]{ulem}
\usepackage[super,sort&compress,comma]{natbib}

\newcommand{\beq}{\begin{equation}}
\newcommand{\eeq}{\end{equation}}
\newcommand{\dbar}{\overline{d}}
\newcommand{\rbar}{\overline{r}}
\newcommand{\rhol}{\rho_l}
\newcommand{\rhof}{\rho_f}
\newcommand{\drho}{\Delta\rho}
\newcommand{\rhoo}{\rho_{\mathrm{oil}}}
\newcommand{\rhow}{\rho_{\mathrm{water}}}
\newcommand{\rhos}{\rho_{\mathrm{PS}}}
\newcommand{\hstar}{h_\ast}
\newcommand{\lc}{\ell_c}
\newcommand{\Ctwo}{C^2}
\newcommand{\deqm}{d_{\mathrm{eqm}}}
\newcommand{\upd}{\mathrm{d}}
\newcommand{\Udipole}{U_{\mathrm{dipole}}}
\newcommand{\Upart}{U_{\mathrm{particles}}}
\newcommand{\Uliq}{U_{\mathrm{liquid}}}
\newcommand{\pmag}{\vec{p}_{\mathrm{mag}}}
\newcommand{\pelec}{\vec{p}_{\mathrm{elec}}}

\begin{document}
\title{\textsf{\textbf{Self-assembly of repulsive interfacial particles via collective sinking}}}
\author{ \textsf{Duck-Gyu Lee$^1$, Pietro Cicuta$^2$ and Dominic Vella$^1$}\\ 
{\it$^1$~Mathematical Institute, Woodstock Rd, Oxford, OX2 6GG, UK}\\
{\it$^2$~BSS, Cavendish Laboratory, Cambridge, CB3 0HE, UK}
}

\date{\today}
\maketitle
\hrule\vskip 6pt

\begin{abstract}
Charged colloidal particles trapped at an air--water interface are well known to form an ordered crystal, stabilized by a long ranged repulsion; the details of this repulsion remain something of a mystery, but all experiments performed to date have confirmed a dipolar-repulsion, at least at dilute concentrations. More complex arrangements are often observed, especially at higher concentration, and these seem to be incompatible with a purely repulsive potential. In addition to electrostatic repulsion, interfacial particles may also interact via deformation of the surface: so-called capillary effects. Pair-wise capillary interactions are well understood, and are known to be too small (for these colloidal particles) to overcome thermal effects. Here we show that collective effects may significantly modify the simple pair-wise interactions and become important at higher density, though we remain well below close packing throughout.  In particular, we show that the interaction of many interfacial particles can cause much larger interfacial deformations than do isolated particles, and show that the energy of interaction per particle due to this ``collective sinking'' grows as the number of interacting particles grows. Though some of the parameters in our simple model are unknown, the scaling behaviour is entirely consistent with experimental data, strongly indicating that estimating interaction energy based solely on pair-wise potentials may be too simplistic for surface particle layers.
\end{abstract}
\vskip 6pt
\hrule

\section{Introduction}

Since first being observed by Pieranski in 1980 \cite{Pieranski1980}, the self-assembly of colloidal particles at a liquid--fluid interface has sparked considerable interest as both a system in which to study the nature of phase transitions in two-dimensions \cite{Kalia1981,Hurd1985} and as a useful tool for designing regular arrays at small scales \cite{Isa2010}. A particularly striking feature of these self-assembled colloidal crystals is that the constituent colloids maintain an equilibrium separation that can be several times larger than the particle diameter \cite{Parolini2015}: this feature is particularly useful in particle lithography  since it allows the fabrication of patterns without particle--particle contact, which can cause  cross-talk in various applications\cite{Isa2010}. At a fundamental level, this separation clearly indicates that the particles are repulsive; detailed studies show that the charge on the particles is, in fact, separated so that the colloids  behave as electrical dipoles\cite{Aveyard2002,Masschaele2011} or even quadrupoles \cite{Boneva2007,Boneva2009}.

Pieranski\cite{Pieranski1980} believed the interaction between particles to be purely repulsive: he attributed the regular spacing of colloids to be due to the geometrical confinement of the system. However, isolated clusters of particles have been observed experimentally\cite{Ghezzi1997,RuizGarcia1997,RuizGarcia1998,Stamou2000}, suggesting that this geometrical confinement is not necessary for the formation of a well-ordered crystal. To form such a bound state requires a long-ranged attractive force to overcome the electrostatic repulsion. The existence of such an attraction was also suggested from the inversion of the pair-correlation function\cite{Quesada2001} and could be at the origin of stable clusters studied by Nikolaides \emph{et al.}~\cite{Nikolaides2002}. However, the question of what provides the attractive interaction that balances this repulsion remains open.

At first sight, it is natural to assume that this attraction could result from the well-known attraction between floating particles that is mediated by meniscus deformation \cite{Chan1981,Kralchevsky2000,stebe15}, sometimes called the `Cheerios effect' \cite{Vella2005}. However, it is also well-known that for pairwise interactions, the interaction energy of these flotation forces\cite{Kralchevsky2000} becomes on the order of the thermal energy, $k_BT$, for particles of radius smaller than around $10\mathrm{~\mu m},$\cite{Kralchevsky2000} which is precisely the scale at which these colloidal crystals are observed. Nevertheless, an attractive force persists and so the question remains: what is the basic mechanism behind the attractive force?

Several different mechanisms have been proposed as the origin of the attractive force, including undulations due to a rough contact line \cite{Stamou2000} and/or enhanced normal forces of electrical origins \cite{Nikolaides2002,Nikolaides2003,Boneva2009} --- an electro-dipping force. However, none of these explanations are able to satisfactorily explain all experimental observations: particle roughnesses of the size suggested by Stamou \emph{et al.}\cite{Stamou2000} were not observed experimentally \cite{Bresme2007} while observations with imposed electrical fields to vary the strength of the electro-dipping force did not produce the expected variation in particle spacing\cite{Masschaele2011}. We are not able to resolve these disagreements here, or to propose a new electrostatic mechanism. Instead, we revisit the tacit assumption that  the interaction energy between a pair of particles is a useful way of estimating the typical interaction energy for a large number of interfacial particles.

The qualitative change to the flotation of particles caused by other nearby objects is now well documented \cite{Vella2006,Abkarian2013,Vella2015}. At macroscopic scales, rafts of dense objects float significantly deeper in the liquid than they do in isolation. This is because the proximity of other particles in the raft constrains the menisci to be more horizontal than they would be for an isolated particle: the particles thus sink deeper into the liquid so that hydrostatic pressure can make up for the loss of supporting force from surface tension. Indeed, this effect can be so dramatic that dense particles that are able to float in isolation may actually sink in the proximity of enough other floating particles \cite{Vella2006,Abkarian2013}. While the dramatic loss of floating stability is unlikely to be relevant at the very small scales of colloidal particles, the observation that their vertical force balance may be affected by the presence of other particles is likely to be robust. The question we address in the remainder of this paper is how any alteration to the vertical force balance manifests itself in the horizontal force balance condition --- is the effective interaction energy between particles substantially modified from the two-body case? This question has been addressed previously using mean-field, or coarse-grained approaches \cite{Pergamenshchik2009,Pergamenshchik2012,Dominguez2010,Bleibel2011PRL,Bleibel2011EPJE,Bleibel2014,Bleibel2016}; here, we study this problem by considering in detail the meniscus deformations caused by individual particles and the collective effect of this deformation.

We begin by considering a two-dimensional model problem in \S\ref{sec:2D}, which allows us to obtain numerical results for large assemblies of particles. These results can be understood quantitatively using a scaling analysis, which is then extended to the three-dimensional problem of most interest in \S\ref{sec:3D}. We then conclude in \S\ref{sec:conclusions} by discussing the possible significance of our theoretical results for the spontaneous formation of colloidal islands.

\section{Two-dimensional formulation\label{sec:2D}}

Here we consider a two-dimensional configuration in detail. With this simplification, we are then able to consider a point-like particle (corresponding to a line in 3D), facilitating our analysis. We aim to gain physical understanding here, that can then be used to inform an understanding of the fully three-dimensional problem.

\subsection{Key ingredients of the model}

\begin{figure}
  \centerline{\includegraphics[width=0.95\columnwidth]{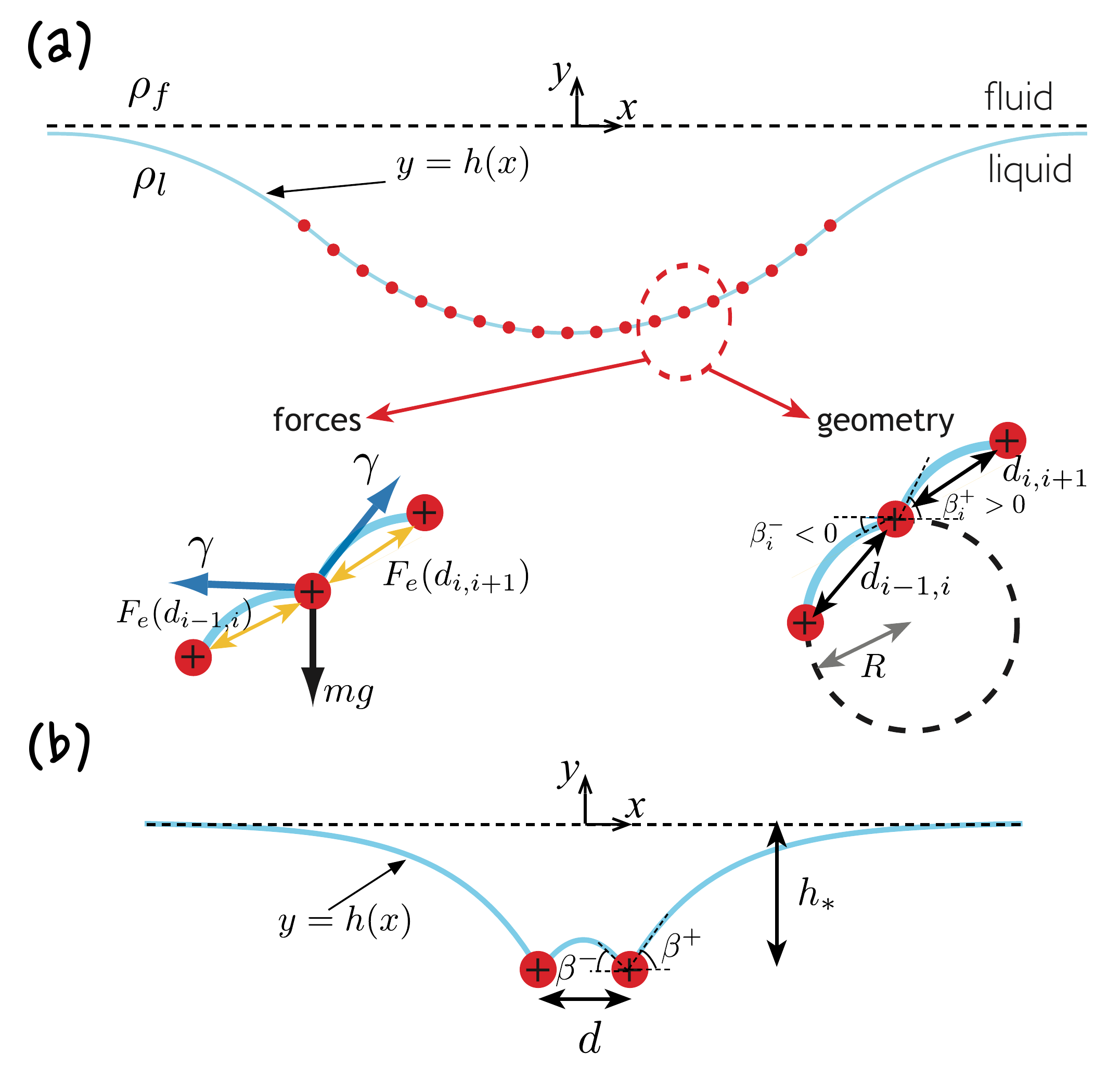}}
  \caption{Geometry of floating line charges at the interface between  a liquid and a fluid. (a) The general formulation of the problem for many charges, showing the geometrical and force balance aspects as insets. (b) The specialization to the two-body problem. Note that, in each case, the angles of the interface to the horizontal are defined positive above the horizontal.}
\label{fig:setup}
\end{figure}

We are interested in understanding in detail the flotation of a series of objects; to be able to form the crystals that are observed, we expect that these objects should be subject to both attractive and repulsive interactions. The question of interest is then really can an ordinarily small pairwise interaction be amplified in a many-body interaction? The simplest purely repulsive interaction in 2D is that between line charges (rather than dipoles, which can become attractive depending on their orientation). By `line charge' here, we mean the two-dimensional analogue of a point charge: charge exists along a line perpendicular to the plane of figure \ref{fig:setup}. We therefore consider a collection of $N$ identical line charges floating at the interface between a liquid, of density $\rhol$, and a fluid, of density $\rhof<\rhol$; the interfacial tension is $\gamma$. Each line charge has an electric charge $+q$ per unit length (so that electrostatic interactions are purely repulsive) as well as a weight per unit length $mg$. A sketch of the setup is shown in figure \ref{fig:setup}.

In this simple model, the point-like particles deform the interface purely due to their weight: there are no wetting effects to be considered. We envisage that this weight-induced deformation will be small since the particles of interest are themselves small and easily supported by surface tension. As a consequence, the attractive interaction due to the weight-induced meniscus deformation between a pair of these particles should also be small, leading to a relatively large equilibrium separation at which capillary attraction balances electrostatic repulsion. Equivalently, we expect the typical interaction energy between a pair of such particles ($N=2$) to be very small (compared to the thermal energy). To understand the balance between deformation-induced attraction and electrostatic repulsion, we first consider the two-body problem in some detail.

\subsection{The two-body problem}

The interaction between two identical line charges will clarify the various dimensionless parameters that control the problem. We orientate our axes so that the two particles are located at $x=\pm d/2$ with $d$ the distance that separates them (see fig.~\ref{fig:setup}b). The meniscus $y=h(x)$ must satisfy the Laplace--Young equation
\beq
\gamma \frac{h''}{\bigl[1+(h')^2\bigr]^{3/2}}=(\rhol-\rhof)gh,
\label{eqn:LaplaceYoung}
\eeq with primes denoting differentiation with respect to $x$. Eqn \eqref{eqn:LaplaceYoung} is to be solved subject to a symmetry condition, $h'(0)=0$, and the meniscus decay condition, $h(\pm\infty)=0$. However, the meniscus may have a discontinuity in slope at each particle (indeed this discontinuity is what leads to a horizontal force between particles in this line-particle limit).

\subsubsection{The linearized problem }

Under the assumption that the particles only deform the interface slightly, so that the slope of interface deformations $h'\ll1$, the Laplace--Young equation \eqref{eqn:LaplaceYoung} may be linearized. The deformation of the interface caused by a single line charge at $x=x_i$ may then be found analytically to be
\beq
\frac{h(x)}{\lc}=-\frac{W}{2}\exp(-|x-x_i|/\lc)
\eeq where $x$ is the horizontal coordinate measured from the mass,
\beq
\lc=\left(\frac{\gamma}{(\rhol-\rhof)g}\right)^{1/2}
\eeq is the capillary length and
\beq
W=mg/\gamma
\label{eqn:Wdefn}
\eeq is the dimensionless weight per unit length of the mass. Here, the prefactor is determined by the vertical force balance on the mass --- the vertical force provided from surface tension must balance the weight of the line mass.

If two identical masses float with separation $d$ then by linear superposition\cite{Nicolson1949}, we have
\begin{align}
\frac{h(x)}{\lc}\approx-W\begin{cases}
\exp(-d/2\lc)\cosh(x/\lc),& |x|<d/2\\
\cosh(d/2\lc)\exp(-|x|/\lc), &|x|>d/2.
\end{cases}
\label{eqn:2bodyMen}
\end{align}

To determine the separation distance at equilibrium, $d$, we use the horizontal balance between the (repulsive) electrostatic force and the (attractive) capillary forces, which reads
\beq
\gamma(\cos\beta^+-\cos\beta^-)+\frac{q^2}{2\pi\epsilon_0 d}=0
\label{eqn:HFbal_nl}
\eeq where $\epsilon_0$ is the permittivity of free space and the angles $\beta^\pm$ are the interfacial inclinations at the contact point, given in terms of the meniscus profile by $\tan\beta^\pm=\pm h'(d_\pm/2)$. Using \eqref{eqn:2bodyMen} to determine the leading order behaviour of $\cos\beta^+-\cos\beta^-$ for $W\ll1$ we find that the equilibrium separation is the solution of the equation
\beq
\frac{\Ctwo}{W^2}=\frac{d}{2\lc}\exp(-d/\lc),
\label{eqn:2bodyD}
\eeq where
\beq
\Ctwo=\frac{q^2}{2\pi\epsilon_0\gamma\lc}
\eeq is the dimensionless charge parameter, which measures the strength of electrostatic repulsion at separation $d=\lc$ in comparison to the typical force from surface tension.

A sketch of the RHS in \eqref{eqn:2bodyD} reveals that it is a non-monotonic function of $d/\lc$; in particular, an equilibrium is only possible for sufficiently weak repulsion, or sufficiently large weight, so that $\Ctwo/W^2\leq 0.184$. For a given value of $\Ctwo/W^2\leq 0.184$ there are two equilibria, the smallest of which is stable and the largest of which is unstable. For $d/\lc\ll1$, the position of the stable equilibrium is given by
\beq
\frac{d}{\lc}\approx 2\frac{\Ctwo}{W^2}.
\label{eqn:sep2Body}
\eeq 

The analytical progress allowed by the assumption of a linear yields key insight. In particular, we see that as $C^2/W^2$ decreases, so does the equilibrium separation between them, $d$. As a result of this, the depth at which each particle floats
\beq
\frac{h_\ast}{\lc}=-\frac{W}{2}\left[1+\exp(-d/\lc)\right],
\eeq \emph{increases} as $C^2/W^2$ \emph{decreases}. This sinking is caused by the presence of nearby objects and so we refer to it as `collective sinking' here. The fact that the presence of nearby objects modifies the vertical force balance and hence can cause objects that would float in isolation to sink, has been observed at macroscopic scales previously \cite{Vella2006,Vella2007,Abkarian2013,Vella2015}. Here, we do not consider this sinking transition, but emphasize the key point that  the presence of a second particle nearby, via its interfacial deformation, modifies the behaviour of a first particle. This  is similar to the capillary collapse studied recently \cite{Dominguez2010,Bleibel2011PRL,Bleibel2011EPJE,Bleibel2014,Bleibel2016}. We shall shortly go beyond the mean-field approach adopted in these previous works by explicitly considering ensembles of particles accounting for the interface deformation beyond the linear theory just presented. To see the possible effect of the nonlinearities, we first reconsider the two-body problem, accounting for nonlinear meniscus deformation.

\subsubsection{The nonlinear problem~}

The above analysis hinged on the assumption that the slope of the interface, $h'\ll1$, so that the Laplace--Young equation \eqref{eqn:LaplaceYoung} could be linearized and the meniscus deformations caused by each particle superposed. While this is a valid assumption for large particle separations and small particle weights, we now investigate how the results of the previous sub-section change once nonlinear interface deformations are considered.

\begin{figure}
  \centerline{\includegraphics[width=0.7\columnwidth]{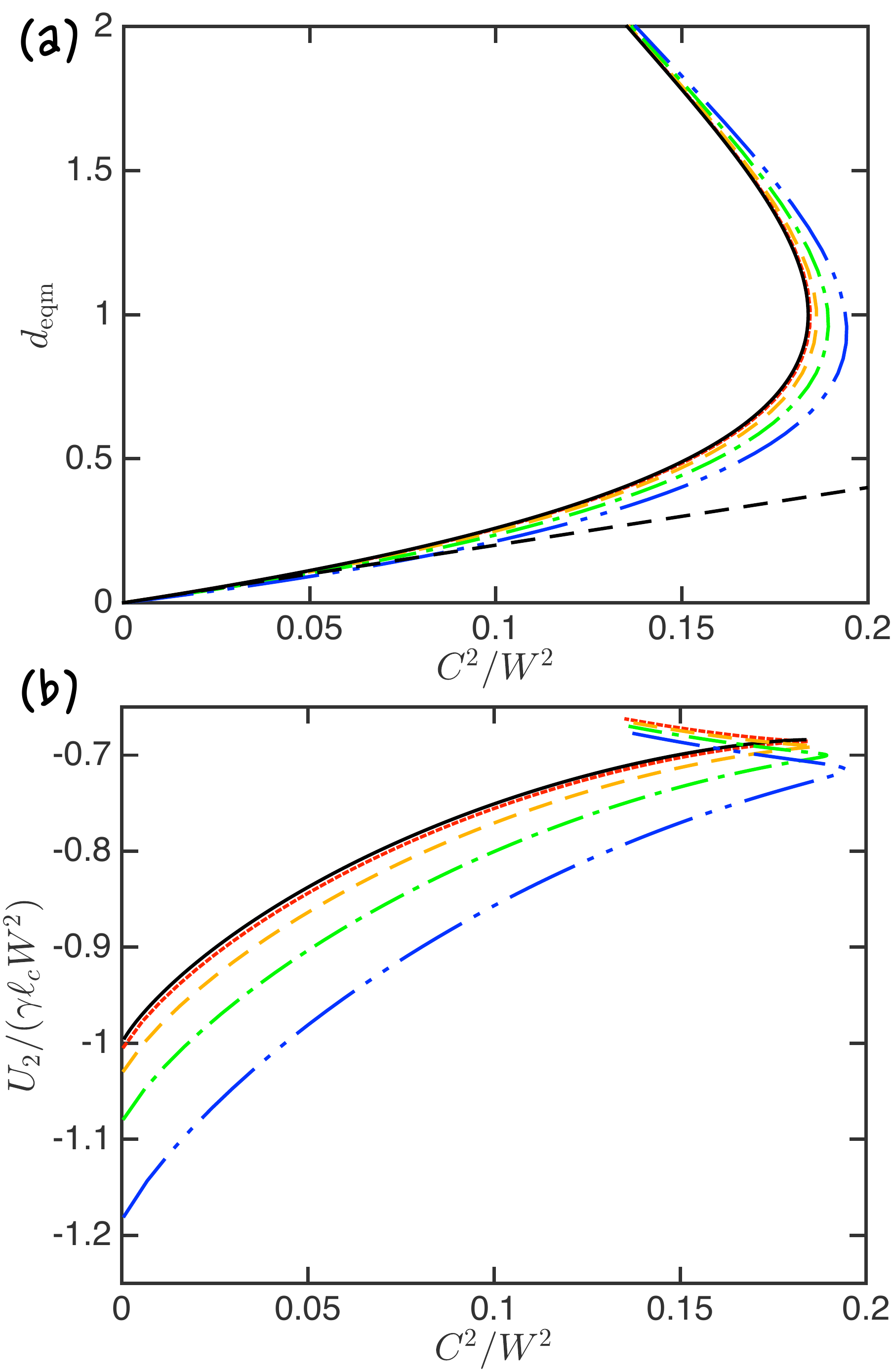}}
  \caption{Comparison between the linearized and fully nonlinear approaches to the two body problem. (a) The equilibrium separation $\deqm$ as a function of the charge-to-weight ratio $C^2/W^2$. The result of the linear analysis \eqref{eqn:2bodyD} (solid black curve) are shown together with the result of full nonlinear computations for $W=0.2$ (red dotted), $W=0.4$ (orange dashed), $W=0.6$ (green dash-dotted) and $W=0.8$ (blue dash-double dotted). The black dashed line represents the asymptotic result \eqref{eqn:sep2Body}, valid for $\Ctwo/W^2\ll1$. (b) The combined energy of the system (compared to that of a flat interface and infinitely separated charges) at the corresponding equilibrium separation, $\deqm$. Here the result of the linear analysis \eqref{eqn:bindEnergy2part} (solid black curve) is shown together with the result of full nonlinear computations for $W=0.2$ (red dotted), $W=0.4$ (orange dashed), $W=0.6$ (green dash-dotted) and $W=0.8$ (blue dash-double dotted). Note that two equilibria exist here, but that one corresponds to a higher energy, and hence is unstable.}
\label{fig:2BodyEnergy}
\end{figure}

When the meniscus slope is no longer considered to be small, the Laplace--Young equation must be solved numerically. In fact, all that is required is to find the angles that the menisci make with the horizontal, $\beta_\pm$ in fig.~\ref{fig:setup}. This simplification, and the fact that the external menisci extend to infinity, mean that we may make use of well known\cite{Mansfield1997} first integrals of the Laplace--Young equation \eqref{eqn:LaplaceYoung} which give
\begin{align}
\sin\beta_+&=-\frac{h_\ast}{\lc}\bigl[1-h_\ast^2/(4\lc^2)\bigr]^{1/2} \label{eqn:LYfirstInta}\\
\cos\beta_+&=1-\frac{h_\ast^2}{2\lc^2}. \label{eqn:LYfirstIntb}
\end{align}
To determine the angle $\beta_-$, however, we must resort to a numerical solution of \eqref{eqn:LaplaceYoung} subject to the boundary conditions
\beq
h'(0)=0,\quad  \sin\beta_-=W+\frac{h_\ast}{\lc}\bigl[1-h_\ast^2/(4\lc^2)\bigr]^{1/2}.
\eeq (These relations express symmetry and vertical force balance, respectively.) Once $\beta_-$ and $h_\ast$ have been determined for a particular configuration, the horizontal force balance \eqref{eqn:HFbal_nl} gives the dimensionless charge $C$ required for flotation at that equilibrium separation. The results of this numerical calculation, and a comparison with the corresponding result for small deformations \eqref{eqn:2bodyD}, are shown in fig.~\ref{fig:2BodyEnergy}a. We observe that the trend is very similar to that observed in the linear theory, although the nonlinear equilibrium separation at fixed $C^2/W^2$ decreases as $W$ increases: the nonlinear effect of nearby particles is to draw those particles closer together than would be supposed from the linear theory.

\subsubsection{The energy of interaction~}

For another perspective on the problem, we consider the energy of the system, $U_2$, which is  given in dimensional terms by
\begin{eqnarray}
\tfrac{1}{2}U_{2}=mg \hstar&-&\frac{q^2}{4\pi\epsilon_0}\log(d/\lc)\nonumber\\
&+&\tfrac{1}{2}\int_0^\infty\left[(\rhol-\rhof)gh^2+\gamma (h')^2\right]~\upd x.
\end{eqnarray} 
For the case of linear deformations, this expression may be evaluated and expressed  in dimensionless terms as
\beq
\frac{U_{2}}{\gamma\lc}=-W^2 \left[\frac{\Ctwo}{W^2}\log(d/\lc)+\tfrac{1}{2}\left(1+e^{-d/\lc}\right)\right].
\label{eqn:bindEnergy2part}
\eeq This expression is compared to the fully nonlinear calculations at the equilibrium separation, $\deqm$, in fig.~\ref{fig:2BodyEnergy}b. From this plot we observe that as the electrical repulsion parameter $\Ctwo$ increases, the depth of the energy well in which the system sits actually decreases: increasing the strength of repulsion decreases the binding that surface tension and gravity are able to supply until ultimately the particles disperse, separating to $d=\infty$. (This unbinding happens for $\Ctwo/W^2\gtrsim 0.184$ in the linear calculation.) We also note that the small deformation (linear) theory is able to give a very good qualitative account of the results of the nonlinear computations provided that the weight per unit length, $W$, does not become too large. However, the general trend is that, once nonlinear deformations are accounted for, the binding energy is larger (since, as we already saw, the equilibrium separation is smaller). 

\section{The 2-D many-body problem}

\subsection{Governing equations}

In the last section, we considered the two-body problem in some detail. This allowed us to identify the important dimensionless parameters as the dimensionless weight per unit length, $W$, and the dimensionless repulsion strength, $\Ctwo$. Furthermore, we showed that in the limit of light particles, $W\ll1$, there is an equilibrium floating arrangement only if the charge-to-weight ratio $\Ctwo/W^2\leq0.184$. Finally, we found that the typical energy scale of the interaction in such an equilibrium floating arrangement is $W^2$: as we expect the energy of interaction is small when the weight of the particles themselves is small.

We now turn to the many-body problem: does the presence of many floating objects cause a raft of charged particles to float deeper in the liquid than would be the case without many-body interactions? If so, how does this `collective sinking' influence the typical energy well in which each particle sits?

We consider the same setup as for the two-body problem but with $N$ line charges, i.e.~$N$ line charges, each of mass $m$ and charge $q$ per unit length  float at a liquid--fluid interface, as shown schematically in figure \ref{fig:setup}. (For simplicity, we shall consider $N=2n+1$ odd, which facilitates our calculations; we do not expect this restriction to have any material effect, especially for large $N$.) The position of each particle in this `raft' is determined by the balance of forces in both the vertical and horizontal directions.

In the \emph{horizontal} direction  force balance requires the net horizontal force from surface  tension on the $i^{\mathrm{th}}$ particle, $\gamma(\cos\beta_i^+-\cos\beta_i^-)$, to balance the horizontal component of the electrical repulsive force arising from every other particle. In dimensionless terms we have
\begin{equation}\label{eqn:hfbal}
\cos\beta_i^+-\cos\beta_i^-=C^2\sum\limits_{j=-n\atop j\neq i}^n\frac{x_j-x_i}{d_{i,j}^2},
\end{equation} where $d_{i,j}=\bigl[(x_j-x_i)^2+(y_j-y_i)^2\bigr]^{1/2}$ is the distance between two particles. (Note that the Coulombic repulsion between line charges $\sim 1/d_{i,j}$ with the additional factor $(x_j-x_i)/d_{ij}$ coming from resolving the force in the horizontal direction.)

In the \emph{vertical} direction, the restoring vertical force from surface tension on the $i^{\mathrm{th}}$ particle, $\gamma(\sin\beta_i^++\sin\beta_i^-)$, must balance both the weight of the particle, $mg$, and any vertical component of the repulsion between them. We have in dimensionless terms
\begin{equation}\label{eqn:vfbal}
\sin\beta_i^++\sin\beta_i^-=C^2\sum\limits_{j=-n\atop j\neq i}^n\frac{y_j-y_i}{d_{i,j}^2}+W.
\end{equation}

The equations representing force balance give $2N$ equations for $4N$ unknowns (for each particle, we  know neither its $(x_i,y_i)$ coordinates nor the meniscus angles on either side of it, $\beta_i^{\pm}$). To determine additional equations, we must also obtain additional relationships for the $\beta_i^\pm$. In principle, these angles may be determined by solving the Laplace--Young equation \eqref{eqn:LaplaceYoung} subject to the boundary conditions $h(x_i)=y_i$, $h(x_{i+1})=y_{i+1}$. In practice, this calculation is made simpler by two observations: firstly, for the menisci that extend to $\pm\infty$ we may use the first integrals \eqref{eqn:LYfirstInta} and \eqref{eqn:LYfirstIntb}. Secondly, for particles that are close (in comparison to the capillary length $\lc$) the meniscus may be approximated by the arc of a circle, with radius of curvature $R$ (see lower right inset of fig.~\ref{fig:setup}a). The radius of curvature of the meniscus between particles $i$ and $i+1$, which we denote $R_{i,i+1}$, is then determined by noting that the hydrostatic pressure within the liquid along the interface, which we estimate as $-(\rhol-\rhof)g(y_i+y_{i+1})/2$, must be balanced by the capillary pressure drop, $\gamma/R_{i,i+1}$; in dimensionless notation we therefore have
\beq
R_{i,i+1}\approx \frac{2}{y_i+y_{i+1}}.
\label{eqn:CircR}
\eeq Elementary geometry then leads to  expressions for $\beta_i^+$ and $\beta_{i+1}^-$ in terms of the particle positions and $R_{i,i+1}$:
\begin{align}
\beta_i^+&=\sin^{-1}\frac{d_{i,i+1}}{2R_{i,i+1}}+\sin^{-1}\frac{y_{i+1}-y_i}{d_{i,i+1}}\label{eqn:geom1}\\
\beta_{i+1}^-&=\sin^{-1}\frac{d_{i,i+1}}{2R_{i,i+1}}-\sin^{-1}\frac{y_{i+1}-y_i}{d_{i,i+1}}\label{eqn:geom2}.
\end{align}

With the geometrical relationships \eqref{eqn:geom1}--\eqref{eqn:geom2}, the pertinent results from the Laplace--Young equation \eqref{eqn:LYfirstInta}, \eqref{eqn:LYfirstIntb} and \eqref{eqn:CircR} and the two force balance equations \eqref{eqn:hfbal}--\eqref{eqn:vfbal}, we have a closed problem. We solve these equations numerically using Newton's method: an initial guess for the position of the particles ($x_i$, $y_i$) is supplied, which is refined by the iteration step until a convergence criteria for ($x_i$, $y_i$) is met. More details about the numerical method are discussed in Appendix A.

\subsection{Numerical Results}

For rafts consisting of identical particles, there are three parameters that characterize the raft shape: the dimensionless weight per unit length, $W$, the dimensionless charge per unit length, $C$, and the number of particles in the raft, $N$. For given values of these parameters, the theoretical formulation given in the preceding section allows us to determine numerically the position of each particle and the properties of the raft.

\begin{figure}
\centering
\includegraphics[width=0.95\columnwidth]{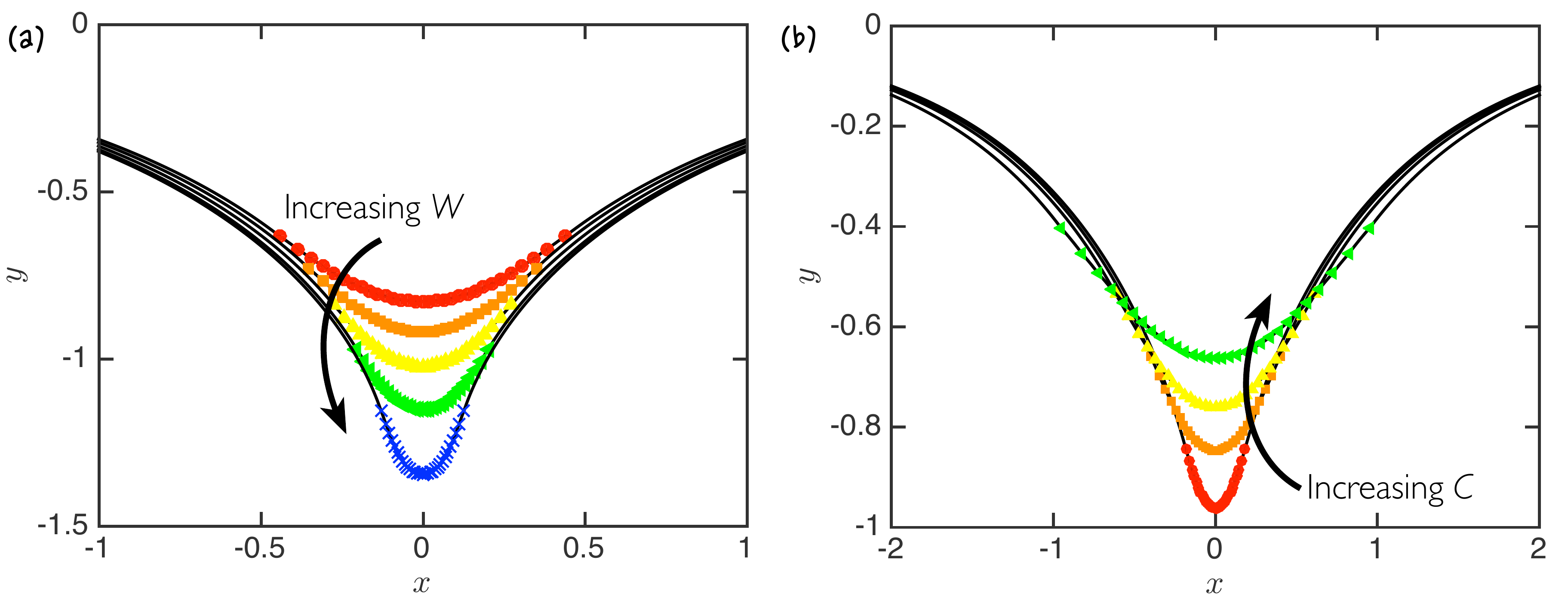}
  \caption{The effect of changing the physical parameters on the shape of a raft with a fixed number of point-like particles  at an interface (here $N=31$). (a) Increasing the weight per unit length of each particle, $W$, causes the raft to sink deeper into the supporting liquid; points show the position of the particles with $C=0.02$ (fixed) and $W=0.0602$ ($\bullet$), $W=0.063$ ($\blacksquare$), $W=0.066$ ($\blacktriangle$), $W=0.069$ ($\blacktriangleleft$) and $W=0.0714$ ($\times$). (b) Increasing the charge per unit length of each particle, $C$, causes the raft to lift out of the supporting liquid (since the electrical repulsion is stronger, the equilibrium distance is larger and particles can reach equilibrium without sinking so deep into the liquid); points show the position of the particles with $W=0.0602$ (fixed) and $C=0.0155$ ($\bullet$), $C=0.0195$ ($\blacksquare$), $C=0.0214$ ($\blacktriangle$) and $C=0.0228$ ($\blacktriangleleft$). In each case, the interface shape is shown by the solid black curves.}
\label{fig:NumsChangePars}
\end{figure}

Our main interest lies in the effect of varying the number of particles $N$ in a raft, and in understanding how a large number of particles behave collectively. However, it is also of interest to see how, with a fixed number of particles, a raft behaves as the two physical parameters, namely $W$ and $C$, are changed. Figure \ref{fig:NumsChangePars}a shows how the raft shape changes as $W$ increases. As should be expected, the particles fall deeper into the supporting liquid as they grow heavier, becoming more closely packed as they do so. However, we emphasize that this process is highly nonlinear: the largest $W$ used in fig.~\ref{fig:NumsChangePars}a is within $20\%$ of the smallest value and yet the maximum depth of the raft increases by almost a factor of $2$ and the particles come significantly closer together. This nonlinearity is a result of the `collective sinking' of the particles: an increase in $W$ brings them closer together, decreasing the vertical supporting force that surface tension is able to provide, causing the particles to lower themselves further into the liquid to achieve that supporting force and, in the process, increasing the attractive force between them.

Figure \ref{fig:NumsChangePars}b confirms the important role of this `collective sinking' in determining the raft shape: as the charge carried by each particle increases, the distance between those particles also increases (since the repulsive force increases). This means that the vertical surface tension force required to support the particles can be obtained at a lower depth and so the raft rises out of the lower liquid.

\begin{figure}
\centering
\includegraphics[width=0.95\columnwidth]{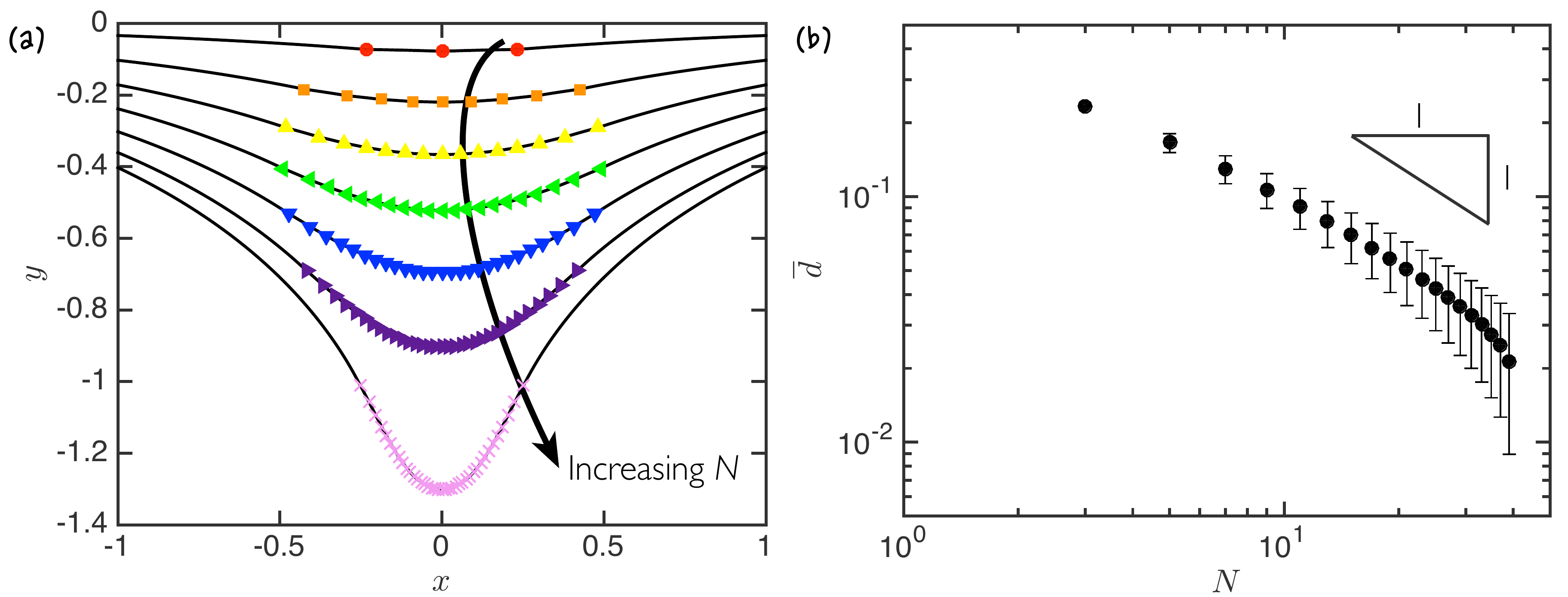}
  \caption{The effect of changing \emph{just} the number of particles at the interface. With $W=0.0602$ and $C=0.02$ fixed, we see that as the raft size grows, particles are not only more closely packed (on average) but also sink lower into the supporting liquid.  (a) Particle positions for rafts with increasing $N$: $N=3$ ($\bullet$), $9$ ($\blacksquare$), $15$ ($\blacktriangle$), $21$ ($\blacktriangleleft$), $27$ ($\blacktriangledown$), $33$ ($\blacktriangleright$), and $39$ ($\times$). (b) Mean distance between neighbouring particles for rafts with $W=0.0602$ and $C=0.02$. Vertical error bars indicate the standard deviation.}
\label{fig:NumsChangeN}
\end{figure}

To understand better the role of collective sinking, figure \ref{fig:NumsChangeN} shows the effect of changing just the number of particles contained in the raft. For very small rafts, $N=3$ for example, the interface is barely deformed, and the equilibrium particle separation is relatively large: this is to be expected since the weight per unit length used here, $W=0.0602\ll1$, does not lead to a significant lateral capillary force. However, as more of these lightweight, lightly charged, particles are introduced (i.e.~$N$ increases), the particles float significantly lower in the liquid (fig.~\ref{fig:NumsChangeN}a) and come much closer together (the mean separation between neighbours, $\dbar$, decreases, as shown in fig.~\ref{fig:NumsChangeN}b). We see then that the attractive capillary interaction between neighbours must be becoming stronger with increasing $N$, since the repulsive electrostatic interaction between neighbours increases as $\dbar$ decreases.

We note that we are not able to find equilibrium cluster shapes with arbitrary values of the weight per unit length $W$ or number of particles, $N$, in a cluster. In particular, for large, heavy clusters ($N$ and $W$ both large) our algorithm fails to find equilibrium configurations. We interpret this apparent lack of equilibrium solutions as a transition from floating to sinking, as has been observed at macroscopic scales with sufficiently large, heavy particle rafts \cite{Vella2006,Abkarian2013,Vella2015}. While this is interesting at a macroscopic scale, we do not study this transition here since this is extremely unlikely to be pertinent at microscopic scales. Instead we focus on how the lowering of the cluster in the liquid (but without becoming immersed in the bulk) modifies the interaction between floating particles. To understand how this `collective sinking' can enhance the strength of lateral capillary interactions, we  turn to some scaling considerations.

\subsection{Scaling analysis and typical energy of interaction}

\begin{figure}
\centering
\includegraphics[width=0.95\columnwidth]{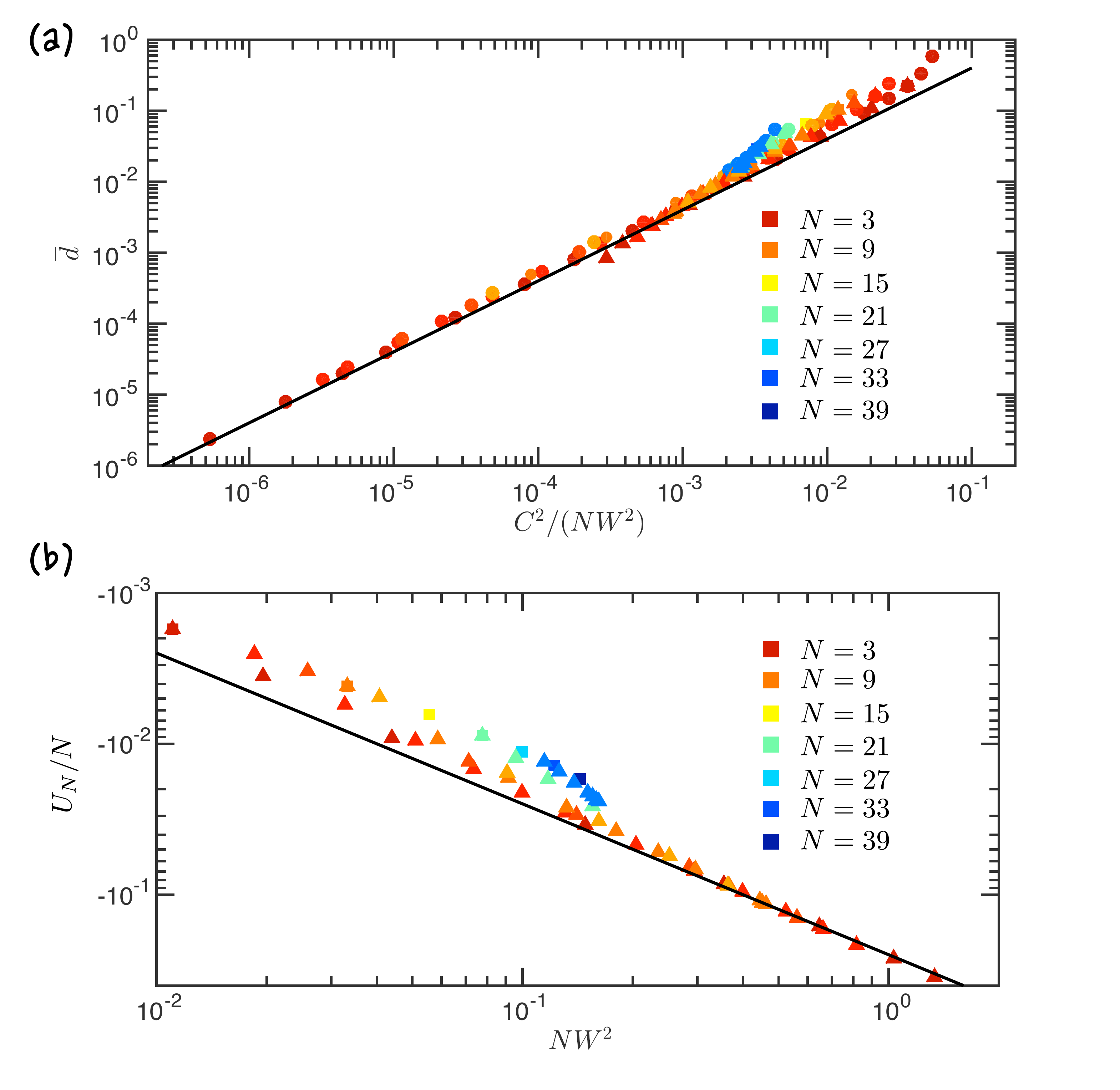}
  \caption{The (a) mean particle separation and (b) energy change per particle as a result of collective sinking. Results are shown for rafts with different numbers of particles, $N$, and different weights per unit length, $W$. Colour is used to show the number of particles in the raft from dark red ($N=3$) to blue ($N=39$) while the symbol indicates different values of $W$: squares show numerical results with $W=0.0602$ and $N$ varying while triangles show individual values of $N$ (coded by colour) with $W$ varying in the range $0.0602\leq W\leq0.66$. Here $C=0.02$ throughout and the solid lines represent the   predictions (a) \eqref{eqn:2Ddbar} and (b) \eqref{eqn:2DbindPrefactor}, which are based on our scaling analysis and comparison with the two-particle problem.}
\label{fig:collapse}
\end{figure}

To understand how collective effects can enhance lateral interactions, we study the total energy of the system in scaling terms. This energy consists of gravitational energy (of displaced liquid \emph{and} particles), interfacial energy and the electrostatic energy of the particles. In scaling terms, the lateral extent of the raft $L\sim \dbar N$; as found in the simulations presented in figs~\ref{fig:NumsChangePars} and \ref{fig:NumsChangeN} we shall assume that $L\ll1$ (i.e.~that the raft is small compared to the capillary length). The gravitational and interfacial energy of the liquid displaced by the raft itself and the outer meniscus $\sim H^2(1+N\dbar)$, where $H$ is the depth of the edge of the raft and `$\sim$' means ``scales as". The gravitational energy of the particles themselves $\sim N W H$. Finally, the electrical potential energy $\sim -C^2N^2\log \dbar$ since there are $N$ particles, each of which interacts with $N-1$ other particles (and where we neglect terms like $N^2\log N$ since they do not vary with $\dbar$). The total energy is then
\beq
U_N\sim H^2(1+N\dbar)+WNH-C^2N^2\log \dbar,
\eeq which may be minimized by varying $H$ and $\dbar$ simultaneously. This minimization gives that  $\dbar\sim NC^2/H^2$ and $H\sim NW/(1+\overline{d}N)\sim NW$, since $N\dbar\ll1$, so that $\dbar\sim \frac{C^2}{W^2} N^{-1}$. Comparing this to the corresponding result from the two-body problem, \eqref{eqn:sep2Body}, and assuming that the prefactor might be such that this result holds all the way down to $N=2$, we then hypothesize that
\beq
\dbar\approx \frac{C^2}{W^2} \frac{4}{N}.
\label{eqn:2Ddbar}
\eeq This prediction is compared with our numerical data in figure \ref{fig:collapse}a. The comparison shows that numerical data collapse as $C$, $W$ and $N$ are varied independently; furthermore the scaling and, to a certain extent, the prefactor predicted in \eqref{eqn:2Ddbar} are as predicted. (Note that strictly speaking the above scaling analysis applied only to large numbers of particles, $N\gg1$, and so the prefactor in \eqref{eqn:2Ddbar} is meant to be indicative.)

The total energy of interaction of the system in this equilibrium, $U_N$, is also of interest. Using the equilibrium value $\dbar$ from \eqref{eqn:2Ddbar}, we find that
\beq
U_N\sim N^2\bigl[W^2+C^2(1+\log N)\bigr]\sim N^2W^2,
\label{eqn:2DbindEnergy}
\eeq for $C/W\ll1$. The key observation about this energy is that it is quadratic in $N$, which means that the binding energy per particle, $U_N/N\sim N$ increases with the number of particles: the collective sinking of particles into the liquid increases their binding energy. Put another way, an estimate of the binding energy that focuses only on the energy of pairwise interactions is qualitatively incorrect as the size of the raft increases. 

Comparing the scaling in \eqref{eqn:2DbindEnergy} with the exact result for $N=2$, \eqref{eqn:bindEnergy2part}, and assuming that the prefactor is such that the former scaling with $N=2$  reduces to \eqref{eqn:bindEnergy2part} we find that
 \beq
\frac{U_N}{N}\approx\tfrac{1}{4}NW^2.
\label{eqn:2DbindPrefactor}
\eeq Figure~\ref{fig:collapse}b shows a plot of the  binding energy (per particle)  determined numerically as the physical parameters of the system are varied. Again, we see that the data collapse using the scaling suggested by \eqref{eqn:2DbindPrefactor}, and that the binding energy per particle exhibits a similar scaling to that expected from \eqref{eqn:2DbindPrefactor}, which is shown as the solid line in figure \ref{fig:collapse}b. As expected, therefore, we see that collective sinking can cause an increase in the `energy well' in which floating particles find themselves.

We emphasize that the pair-wise calculation, which led to \eqref{eqn:bindEnergy2part}, would predict an energy per particle $\sim W^2$. Collective sinking (and also the linear superposition of capillary collapse\cite{Dominguez2010,Bleibel2011PRL}) leads to an additional multiplicative factor $N$, which clearly becomes more important as the size of the cluster,  $N$, increases. In particular, while the scaling in \eqref{eqn:2DbindEnergy} holds, the binding energy per particle can become arbitrarily large as a result of collective sinking.

\section{The 3-D case\label{sec:3D}}

\subsection{Scaling analysis}

In the three-dimensional case of interacting dipoles that motivated this study it is difficult to perform full numerical calculations: these would involve determining the three-dimensional meniscus shape surrounding many objects and resolving a contact line that is not in general circular, even for floating spheres \cite{CoorayThesis}. While such an investigation remains a possibility for the future, here we focus on the understanding we have gained from our detailed consideration of the two-dimensional problem to understand the 3D problem using scaling arguments.

We consider $N$  dipoles, each of mass $m$. Assuming that the dipoles are aligned by an external field so that they are repulsive, not attractive, the pairwise interaction energy may be written $U\sim A/d^3$, where $A$ is a constant that will depend on the nature of the dipolar interaction, e.g.~$A=\mu_0|\pmag|^2$ for magnetic dipoles or $A=|\pelec|^2/\epsilon_0$ for electrical dipoles. The scaling behaviour of the dipolar energy of this assembly deserves careful discussion: in the 2D case the sum of pairwise interaction energies meant that the total energy scaled like $N^2$. For dipoles in 3D, however, the scaling is more subtle since the energy of an individual dipole surrounded by an infinite, planar cloud of dipoles with mean nearest-neighbour spacing $\dbar$ is found by  summing over the interaction energies of an infinite series of rings of radius $R_i=i \dbar$ ($i=1,2,3,...$), each containing approximately $2\pi i$ other dipoles. We therefore find that $U\lesssim \sum_{i=1}^\infty 2\pi i\times A/(i\dbar)^3= 2\pi \zeta(2)A/\dbar^3 $, and  that the energy of the system due to these dipolar interactions is $\Udipole\sim NA/\dbar^3$.  The gravitational energy of the particles is $\Upart\sim Nmg H$, while the gravitational (and interfacial) energy of the liquid due to the deformation is $\Uliq\sim \drho g H^2 \bigl[N^{1/2}\dbar\lc+N\dbar^2\bigr]$, where we have included the displaced liquid from the aggregate itself as well as the external meniscus around the perimeter and $\drho=\rhol-\rhof$. 

Minimizing over $H$, we find that
\beq
H\sim- \frac{Nm}{\drho\bigl[N^{1/2}\dbar\lc+N\dbar^2\bigr]}
\label{eqn:Hmin}
\eeq while minimizing over $\dbar$ gives
\beq
\frac{NA}{\dbar^4}\sim \drho g H^2 \bigl[N^{1/2}\lc+N\dbar\bigr]\sim \frac{N^{3/2}m^2g}{\drho\dbar^2\bigl[\lc+N^{1/2}\dbar\bigr]}.
\label{eqn:3D:deqn}
\eeq Solving for $\dbar$ gives
\beq
\dbar\approx\alpha N^{-1/4}\frac{A^{1/2}\lc^{1/2}\drho^{1/2}}{mg^{1/2}},
\label{eqn:3Ddipole}
\eeq assuming that $N^{1/2}\dbar\ll\lc$ and introducing an unknown prefactor $\alpha$. Note that as in the 2D monopole case, the mean separation decreases as the aggregate grows larger.

We emphasize that this result only holds for large clusters, where each dipole effectively has infinitely many other dipoles with which it could interact; the interaction energy is then cut off by the decay of the dipolar potential, rather than the number of neighbours. With smaller clusters, the dipole--dipole interaction energy is instead limited by the number of available dipoles, which are a typical distance $\rbar$ away. In this case, $\Udipole\sim AN^2/\rbar^3$. Assuming that $\rbar\sim N^{1/2}\dbar$, we have that $\Udipole\sim AN^{1/2}/\dbar^3$. To progress further, we assume that small clusters are approximately spherical\cite{Jones2016}, with radius of curvature $\sim\rbar$ so that the surface energy $\sim \gamma N\dbar^2$; equating with the dipole--dipole energy, we find that $\dbar\sim N^{-1/10}$.

\subsection{Macroscopic analogue experiments}

We are not aware of experimental data at a microscopic scale that would allow  the scaling law in \eqref{eqn:3Ddipole} to be tested. However, recent experiments on macroscopic paramagnetic spheres floating at an air--water interface showed that these spheres do form a raft of the form considered in this paper \cite{Vandewalle2013}. By digitizing the images presented in figure 2 of Vandewalle \emph{et al.}\cite{Vandewalle2013}, we were able to compute the mean particle separation (measured only between nearest neighbours) from these experiments on aggregates with varying numbers of particles, $N$. We expect to see the particle separation decreasing with increasing $N$, and more specifically, according to \eqref{eqn:3Ddipole}, that $\dbar\sim N^{-1/4}$. This scaling  is confirmed by  the results presented in figure \ref{fig:Microscopic}b with the prefactor for this scaling corresponding to  $\alpha\approx3$. We also note that while for small cluster sizes, $N\lesssim 40$, the data presented in fig.~\ref{fig:Microscopic}b appear to flatten out slightly, this is not as much as might be expected on the basis of the $\dbar\sim N^{-1/10}$ scaling discussed above for this limit.

\subsection{Relevance to microscopic scenarios}

 \begin{figure}
  \centering
\includegraphics[width=0.99\columnwidth]{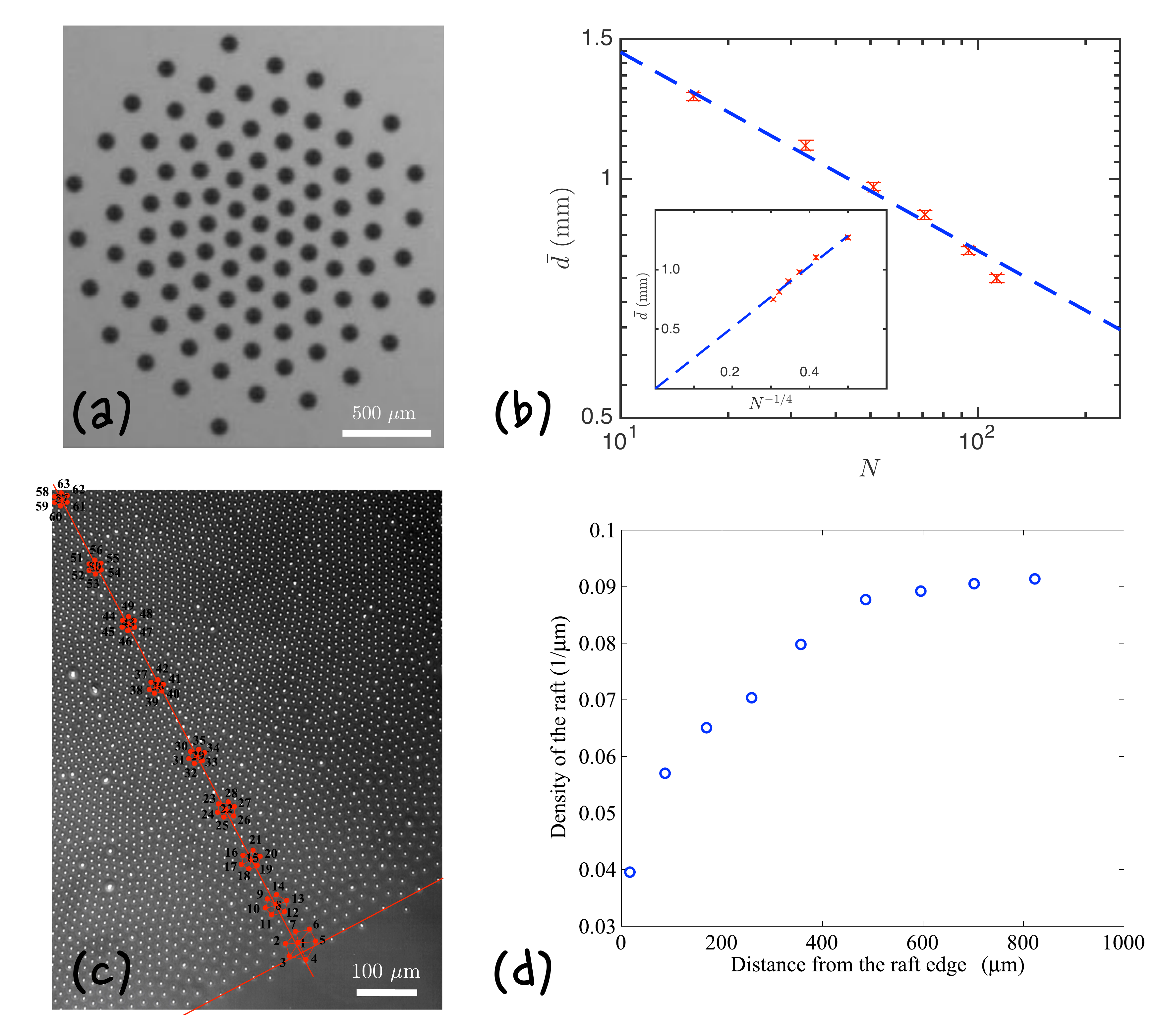}
  \caption{  Aggregates of dipolar particles at a liquid--fluid interface. (a) $N=94$ paramagnetic spheres of radius $a=200\mathrm{~\mu m}$ in an externally applied field of $50\mathrm{~G}$ form a closed aggregate at an air--water interface (image taken from Vandewalle \emph{et al.}\cite{Vandewalle2013} with permission from Springer). Note in particular that the particles near the edge are more  spaced than those at the centre where, since the interfacial deflection is larger, the capillary attraction is larger too. (b) The average distance $\dbar$ between paramagnetic particles floating at the air--water interface (data taken from Vandewalle \emph{et al.}\cite{Vandewalle2013}). The dashed line shows the best fit from the prediction \eqref{eqn:3Ddipole} --- a scaling that is tested further in the inset. (c) PS particles of radius $a=1\mathrm{~\mu m}$ at the interface between decane and a $0.1\mathrm{~M}$ aqueous NaCl solution. Again, we note that near the edge of the cluster the particle spacing becomes larger than it is away from the edge --- an observation that is quantified in (d) by plotting the variation of particle density along a normal to the edge of the aggregate (highlighted by the red lines in (c)).}
\label{fig:Microscopic}
\end{figure}

Having seen that the scaling law in \eqref{eqn:3Ddipole} is able to predict the increased clustering of macroscopic dipolar particles as the size of the aggregate increases, we now extrapolate this scaling law to the microscopic scale that motivated this work: colloidal particles at a liquid--fluid interface. In this setting, the key question is how the typical energy of interaction per particle,
\beq
\frac{\Udipole}{N}\sim \frac{A}{\dbar^3}\sim \frac{N^{3/4}}{\alpha^3}\frac{1}{A^{1/2}} \frac{m^3g^{3/2}}{\lc^{3/2}\drho^{3/2}},
\eeq compares with the thermal energy, $k_BT$. Based on this scaling law, we see that for this particle-level interaction energy to be larger than the thermal energy, we must have $N\gtrsim N_c$ where
\beq
N_c\sim  \alpha^4(k_BT)^{4/3}A^{2/3} \frac{\lc^{2}\drho^{2}}{m^4g^{2}}.
\label{eqn:Ncrit}
\eeq

To make further progress, we need to estimate the size of the various terms in \eqref{eqn:Ncrit}. From detailed studies of the pair-correlation function for  colloidal layers of PS particles (of radius $a=1\mathrm{~\mu m}$ and density $\rhos=1050\mathrm{~kgm^{-3}}$) it has been suggested\cite{Parolini2015} that $A\approx 3\times 10^5k_BT\mathrm{~\mu m}^{3}$. It is very difficult to be certain of the value of the constant of proportionality $\alpha$ from the analogue magnetic experiment (which is likely to depend on, among other things, the wetting properties of the particles), and we see that the scalings above vary sensitively with $\alpha$. Therefore we take $0.1\lesssim\alpha\lesssim1$ for now, and note the typical ranges of the parameters based on this. We also note that the real experiments of interest occur at an oil--water interface with $\rhoo=730\mathrm{~kg m^{-3}}$, $\rhow=1000\mathrm{~kgm^{-3}}$ and $\gamma=52\mathrm{~mNm^{-1}}$ (see Zeppieri \emph{et al.} \cite{Zeppieri2001}, for example); as such, we expect that the driving mass will be $m=\tfrac{4\pi}{3}(\rhos-\rhoo)a^3$ while the appropriate capillary length $\lc=[\gamma/\drho g]^{1/2}$ with $\drho=\rhow-\rhoo\approx270\mathrm{~kgm^{-3}}$.

Using the values above, we find  with $\alpha=0.1$ that $N_c\approx3\times 10^4$, i.e.~an aggregate around $100$ particles in each direction should be stable to thermal noise. While large, this number of particles is not infeasible. If instead $\alpha=1$ then the critical number of particles in an aggregate is $N_c\approx3\times 10^8$, which is so large as to be very difficult to observe.

Another test of the scaling laws is the values predicted for the two spatial scales of the aggregate: the mean inter-particle separation, $\dbar$, and the typical depth of sinking, $H$. By construction, the value of $\dbar$ at $N=N_c$ is $\dbar\sim (A/k_BT)^{1/3}\sim 70\mathrm{~\mu m}$; this is the distance at which the typical electrostatic interaction becomes on the same order as the thermal energy, and so in real aggregates the particle separation is likely to be considerably smaller. More interesting is the prediction from \eqref{eqn:Hmin} that around $N=N_c\approx3\times10^4$ the depth of the aggregate $H\sim 3\mathrm{~nm}$ (using the prefactor $\alpha=0.1$ in \eqref{eqn:3Ddipole}); with $\alpha=1$ we find $H\sim 300\mathrm{~nm}$. These depths are significantly smaller than the $O(10\mathrm{~\mu m})$ depths predicted from a previous mean-field model\cite{Pergamenshchik2009} and, as yet, not detected. The cluster depths we predict  are too small to be directly imaged in microscopy, but should be amenable to optical interferometry or FreSCa cryo-SEM \cite{Coertjens2014}. 

While the importance of collective sinking in aggregates of interfacial colloids remains purely speculative, we can compare the phenomenology of our own experiments with what would be expected on the basis of the collective sinking hypothesis. In particular, the collective sinking hypothesis suggests that isolated clusters of colloids can form and, further, that when they do the particles near the edge of the cluster/aggregate should be more widely spaced than those near the centre of the aggregate. (This is observed both in our numerical simulations, see for example fig.~\ref{fig:NumsChangeN}a, and in experiments on macroscopic paramagnetic particles floating at the interface \cite{Vandewalle2013}, fig.~\ref{fig:Microscopic}a.) Similarly, we are able to see a similar phenomenology in aggregates of PS particles trapped at the interface between decane and aqueous salt solutions (see fig.~\ref{fig:Microscopic}c). These experiments follow the methods of Parolini \emph{et al.}~\cite{Parolini2015} for purification of reagents. After long periods of equilibration (sometimes overnight or even after a few days) regions of crystalline arrangements coexist with completely empty regions see fig.~\ref{fig:Microscopic}c. Here, we highlight the clear edge of the cluster (the red line in fig.~\ref{fig:Microscopic}c) and plot the variation of density with distance normal to this interface at isolated points (highlighted in fig.~\ref{fig:Microscopic}c). This analysis reveals (fig.~\ref{fig:Microscopic}d) that there is a more than two-fold decrease in particle density from the bulk of the crystal to the edge. We are not aware of any other explanation for either the existence of a well-defined edge of a cluster or for this spacing, and will explore this phenomenon in detail in a separate work. 

\section{Conclusions\label{sec:conclusions}}

In this paper, we have presented a toy model of the interaction between repulsive particles at an interface. This model allowed us to consider the interaction of large numbers of particles at an interface and to show that as the number of particles increases the particles actually become more closely bound together. This effect is due to the collective sinking of the particles into the liquid: the proximity of other interfacial particles means that the interface is less curved locally than it would otherwise be and so particles sink lower into the liquid. This in turn increases the magnitude of the attractive interaction between them; while this is qualitatively similar to the capillary collapse studied previously, our detailed calculations with two particles showed that this collective sinking provides a binding energy that is quantitatively stronger than  would be predicted by using a linear superposition argument\cite{Dominguez2010,Bleibel2011PRL}.  Crucially, we expect that the importance of this collective sinking, and the additional binding provided by it, increases as the size of the cluster increases.


We presented detailed numerical results for the flotation of line charges. This allowed us to readily perform detailed numerical simulations of the problem, and to gain understanding that could be translated into a scaling argument and thence into scaling arguments for the problem of several dipolar spheres interacting. To make our models more realistic would require detailed simulations of the meniscus around an array of spherical particles. While this would be an involved procedure, we believe that it may soon be feasible computationally\cite{Cooray2012,CoorayThesis} and, further, may yield new insight beyond existing mean-field theories\cite{Pergamenshchik2009,Pergamenshchik2012,Dominguez2010,Bleibel2011PRL}. In particular, these mean-field theories use a linear superposition of the far-field, small deflection meniscus around an axisymmetric object, $h(r)\sim K_0(r/\lc)$, even though close to small axisymmetric objects a subtly different meniscus form is more appropriate \cite{Lo1983}. This subtlety arises from a balance between the nonlinear curvature terms and suggests further that the linear superposition approach may not be valid here, particularly when the particles approach one another on a scale comparable to their radius. At still closer approach, the effects of the particle roughness may become important\cite{Stamou2000}; we do not expect roughness to play a major role, however, since clustering has been observed with particles that are well-separated compared to the particle roughness scale \cite{Bresme2007}. 

For simplicity, our model did not include the electro-dipping force that is believed to be important in at least some observations of colloidal self-assembly.  As a result, our theoretical predictions are unlikely to be directly applicable to colloidal self-assembly. Nevertheless, the mechanism that we have investigated here should be important regardless of what causes the force normal to the interface. In particular, while the gravitational contribution discussed here may not be dominant in all situations of interest,  a similar effect will exist with an electro-dipping force. We hope that our calculations and scaling arguments will motivate further detailed study of this possibility.

\subsection*{Acknowledgments}

We acknowledge many discussions with L.~Parolini and useful comments on an earlier draft of this paper from Martin Oettel. This research was supported by Basic Science Research Program through the National Research Foundation of Korea (NRF) funded by the Ministry of Education, Science and Technology (NRF-2012R1A6A3A03039558) and the  Korea Institute of Machinery and Materials (KIMM) under Grant NK196D.

\section*{Appendix A: Numerical method}


 To solve the equations of vertical and horizontal force balance, \eqref{eqn:hfbal} and \eqref{eqn:vfbal}, we used Newton's method. We firstly arrange the equations to take the form: $F(X)=[f_1,f_2,\ldots,f_{\frac{N+1}{2}},g_1,g_2,\ldots,g_{\frac{N+1}{2}}]^T,$
where $X=[x_1,x_2,\ldots,x_{\frac{N+1}{2}},y_1,y_2,\ldots,y_{\frac{N+1}{2}}]^T$ is the vector of particle positions for half of the raft (using symmetry). The set of particle positions, $X_\ast$, that solves  the vector function $F(X_\ast)=0$ is obtained by starting from an initial guess, $X^{(0)}$ and repeating the iteration scheme
\beq
X^{(n+1)}=X^{(n)}-J^{-1}F(X^{(n)})
\eeq where $J$ is the Jacobian matrix of $F(X)$, i.e.~$J_{ij}=\partial F_i/\partial x_j$. This iteration continues until the maximal element of $F(X^{(n)})$ (in absolute terms) is below some residual, which we set to be $\epsilon=10^{-13}$ here.

\begin{small}
\footnotesize{

\providecommand*{\mcitethebibliography}{\thebibliography}
\csname @ifundefined\endcsname{endmcitethebibliography}
{\let\endmcitethebibliography\endthebibliography}{}

}



\end{small}

\end{document}